\documentclass{article}

\usepackage[final]{neurips_2025_ml4ps}

\usepackage[utf8]{inputenc} 
\usepackage[T1]{fontenc}    
\usepackage{hyperref}       
\usepackage{url}            
\usepackage{booktabs}       
\usepackage{amsfonts}       
\usepackage{nicefrac}       
\usepackage{microtype}      
\usepackage{xcolor}         
\usepackage{graphicx}       
\usepackage{float}          
\usepackage{amsmath}        

\title{Autonomous Pressure Control in MuVacAS via Deep Reinforcement Learning and Deep Learning Surrogate Models}

\author{%
  Guillermo Rodriguez-Llorente\thanks{Corresponding author.} \\
  HI Iberia\\ Juan Hurtado de Mendoza 14\\ 28036 Madrid, Spain.\\
  \texttt{grodriguez@hi-iberia.es}\\
  Department of Mathematics,\\
  Universidad Carlos III de Madrid,\\ 
  Gregorio Millán Barbany Institute,\\
  Universidad Carlos III de Madrid\\28911 Leganés, Madrid, Spain \\
   \texttt{guirodri@inst.uc3m.es} 
  \And
  Galo Gallardo Romero \\
  HI Iberia\\ Juan Hurtado de Mendoza 14\\ 28036 Madrid, Spain \\
  \texttt{ggallardo@hi-iberia.es}
  \And
  Rodrigo Morant Navascués \\
  HI Iberia\\ Juan Hurtado de Mendoza 14\\ 28036 Madrid, Spain\\
  \texttt{rmorant@hi-iberia.es}
  \And
  Nikita Khvatkin Petrovsky \\
  HI Iberia\\ Juan Hurtado de Mendoza 14\\ 28036 Madrid, Spain \\
  \texttt{nkhvatkin@hi-iberia.es}
  \And
  Anderson Sabogal  \\
  IFMIF-DONES Spain\\ 18130 Escúzar, Granada Spain \\
  \texttt{anderson.sabogal@ifmif-dones.es}
  \And
  Roberto Gómez-Espinosa Martín\footnotemark[1] \\
  HI Iberia\\ Juan Hurtado de Mendoza 14\\ 28036 Madrid, Spain\\
  \texttt{robertogemartin@hi-iberia.es}
}

\begin{document}

\maketitle

\begin{abstract}
The development of  nuclear fusion requires materials that can withstand extreme conditions. The IFMIF-DONES facility, a high-power particle accelerator, is being designed to qualify these materials. A critical testbed for its development is the MuVacAS prototype, which replicates the final segment of the accelerator beamline. Precise regulation of argon gas pressure within its ultra-high vacuum chamber is vital for this task. This work presents a fully data-driven approach for autonomous pressure control. A Deep Learning Surrogate Model, trained on real operational data, emulates the dynamics of the argon injection system. This high-fidelity digital twin then serves as a fast-simulation environment to train a Deep Reinforcement Learning agent. The results demonstrate that the  agent successfully learns a control policy that maintains gas pressure within strict operational limits despite dynamic disturbances. This approach marks a significant step toward the intelligent, autonomous control systems required for the demanding next-generation particle accelerator facilities.
\end{abstract}

\section{Introduction}

Making commercial nuclear fusion energy viable remains a formidable scientific and engineering challenge. A critical step toward this goal is the development of materials capable of withstanding the extreme irradiation conditions inside a reactor. To bridge the gap between current materials science and the requirements of future fusion power plants, the International Fusion Materials Irradiation Facility – Demo Oriented NEutron Source (IFMIF-DONES) is under construction. This unique particle-accelerator-based facility will produce an intense neutron flux by colliding a high-energy deuteron beam with a liquid lithium target, thereby enabling the qualification and testing of candidate materials under prototypical fusion conditions \citep{ifmif-dones-status}. To support the safe and reliable design of IFMIF-DONES, the Multipurpose Vacuum Accident Scenarios (MuVacAS) prototype was developed. MuVacAS serves as a dedicated testbed that physically replicates the final segment of the IFMIF-DONES accelerator beamline, where the deuteron–lithium stripping reaction occurs \citep{sabogal:ipac2023-thpa156-MuVacAS}.

A critical aspect of the MuVacAS operation is the precise regulation of pressure within its ultra-high vacuum system. This is achieved through the controlled injection of argon gas to maintain the specific pressure levels required for the stripping reaction, a task complicated by the system's non-linear dynamics and need to respond to fast, unpredictable disturbances. Traditional control methodologies, such as PID controllers, typically require conservative tuning.  While Deep Reinforcement Learning (DRL) is well-suited for such problems \citep{Degrave2022}, its direct application is slow and risky for the physical equipment. To overcome this barrier, this work presents a data-driven approach: first, a Deep Learning Surrogate Model (DLSM) is trained on real data from the machine to create a safe fast-simulating environment. This environment is then used to train a DRL agent to learn pressure regulation under dynamic disturbances. Although the joint use of these two types of models has been explored in prior studies \citep{rodriguezllorente2024applications-iclr, wang2021surrogate}, such approaches remain rare and have so far been limited to simulation-only contexts. The results from this work demonstrate that the agent trained in this way learns a robust policy that maintains pressure within strict limits, outperforming conventional methods and marking a significant step towards the intelligent, autonomous control required for next-generation scientific facilities.

\section{Methodology}
\subsection{Data acquisition }
\label{sec:daq}
The dataset for this study comprises approximately nine hours of real operational data acquired from the MuVacAS prototype. Measurements were sampled at approximately 1 Hz and include synchronized records of the argon injection rate together with pressure readings from a network of 12 cold-cathode sensors distributed along the vacuum line, from the beam entrance at the beginning of the prototype to the collision chamber at the end, where the deuteron–lithium reaction takes place. Sensors 1 and 2, located near the beam entrance, were excluded from the analysis due to their placement in a low-pressure, high-noise region. The argon injection, positioned near the collision chamber, was dynamically modulated using a random Gaussian field technique (with smoothing parameter $\alpha$ of 2.5 and 3) to create diverse yet smoothly varying training conditions \citep{Lang_2011-GRF}. These profiles, with injection rates ranging from 0 to 1 sccm (standard cubic centimeters per minute), were executed automatically on the physical system via a custom script interfacing with the EPICS control system \citep{osti_10193541-epics}. The resulting pressure measurements, spanning several orders of magnitude from approximately $10^{-6}$ mbar to $10^{-3}$ mbar across the sensor array, were log-transformed for model stability.

\subsection{Deep Learning Surrogate Model }
\label{sec:fno}
The selection of an appropriate surrogate model architecture is critical for creating a high-fidelity digital environment. For this task, the Fourier Neural Operator (FNO) was selected. Unlike traditional neural networks that map between Euclidean spaces, this architecture is specifically designed to learn mappings between infinite-dimensional function spaces .The FNO performs convolution in the frequency domain via the Fast Fourier Transform (FFT), enabling it to efficiently capture long-range spatial dependencies and global correlations \citep{2020arXiv201008895L-FNO}. Formally, this model is trained  with the dataset created in Section \ref{sec:daq} to approximate the gas pressure evolution in time  along the accelerator longitude $z$, as described by the following operator: 
\[
q_t, p_t(z) \xrightarrow{\text{FNO}} p_{t+1}(z),
\]
Where at a given time~$t$, the model takes the  argon injection rate~$q_t$ and the entire spatial pressure distribution~$p_t(z)$, and predicts the resulting pressure distribution at the next time step,~$p_{t+1}(z)$. This single-step, Markovian formulation is a direct requirement for its integration within the DRL training loop. 

The model was implemented and trained using the NVIDIA Modulus framework \footnote{ Now known as NVIDIA Physics NeMo: \url{https://developer.nvidia.com/physicsnemo}} on a final dataset of approximately 30,000 instances (environment transitions with argon injections). The dataset was partitioned into 80\% for training and 20\% for testing, and the model's weights were optimized using the Adam algorithm to minimize the mean squared error between its predictions and the ground-truth pressure distributions from the MuVacAS prototype. See Appendix \ref{ap:fno_hyper} for more details on the model hyperparameters.  Finally, an iterative evaluation of the model given an initial pressure distribution and a set of argon injections through time is displayed in Appendix \ref{ap:fno_eval}. 

\subsection{Deep Reinforcement Learning }
To autonomously regulate the pressure at the lithium target (monitored by the sensor located at the end of the beamline), a DRL agent was trained within a simulated environment constructed around the FNO surrogate model (Section~\ref{sec:fno}) and implemented using the Gymnasium framework \citep{towers2024gymnasiumstandardinterfacereinforcement}. The agent’s objective is to learn an optimal policy function~$\pi$ that adjusts the argon injection at each time step in order to reach the desired pressure in the collision chamber as rapidly as possible and maintain it thereafter. For training, the selected algorithm was  Proximal Policy Optimization (PPO) \citep{schulman2017proximalpolicyoptimizationalgorithms},  already implemented in the Stable-Baselines3 (SB3) library \citep{stable-baselines3}. PPO is an on-policy, gradient-based method that has become a standard in continuous control tasks due to its  training stability, natural handling of continuous action spaces and wide adoption and  proved performance across multiple works. In this algorithm, the policy, i.e., the decision-making mechanism of the agent, is represented by a neural network (see Figure~\ref{fig:network_policy} in Appendix \ref{ap:drl_spaces}). At each step, the agent outputs a single continuous action corresponding to the argon mass flow rate, which directly influences the pressure dynamics.

The observation space provided to the agent consists of 3 values. It contains the log-transformed pressure readings from the sensor located at the collision chamber, the log-transformed pressure objective and the current injection rate. The action space is simply a scalar to control the argon injection rate. The ranges of all variables are given by the values chosen in the dataset creation stage in Section \ref{sec:daq}  and can be seen in detail in Table \ref{tab:env_spaces} of Appendix \ref{ap:drl_spaces}. The agent's behaviour is guided by a single-term exponential reward function that depends only on the absolute difference between the current and target pressures (evaluated in the log-pressure space). Concretely, at time step ~$t$ the reward is:
\[
r_t = \exp\big(-k\,|p_t - p^*|\big),
\]

where $p_t$ is the current (log) pressure at the collision chamber, $p^*$ is the desired objective (log) pressure, and $k=4$ is a sharpness coefficient controlling how quickly the reward decays with error. This function attains its maximum value $r_t=1$ when $p_t=p^*$ and decreases smoothly toward $0$ as the deviation grows; it therefore provides stable, non-negative, and well-conditioned gradients for training. Further details on the agent parameters can be seen in Table \ref{tab:ppo_hparams} of Appendix \ref{ap:drl_spaces}.

\section{Results}

First, the trained DRL agent was systematically evaluated in simulated environments with dynamically assigned pressure targets that were not included in the training set. The corresponding results are presented in Appendix~\ref{ap:drl_eval_sim}. In these evaluations, the agent successfully regulated the argon injection to reach and maintain multiple target pressures. An interesting observation is the asymmetry between upward and downward transitions: achieving lower pressure states generally required more control actions than increasing the pressure, which is consistent with the underlying physical dynamics of the system.

\begin{figure}
  \centering
  \includegraphics[width=1\linewidth]{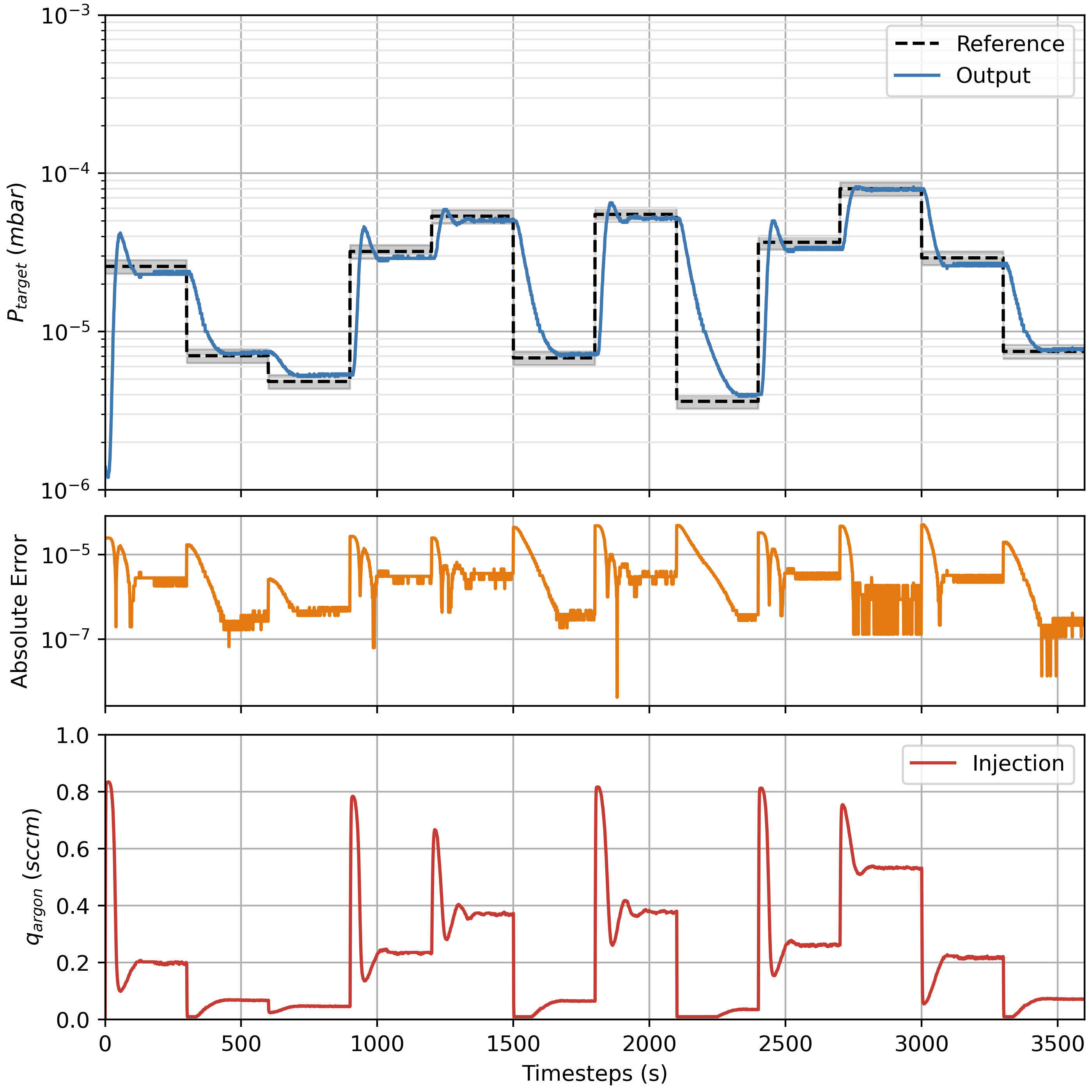}
  \caption{Evaluation of the DRL agent on the real prototype. Top: achieved and reference pressures. Middle: absolute error between the achieved and reference pressures. Bottom: argon injection values.  Here $p_{target}$ denotes the pressure in the lattice section where the collision chamber is located.}
  \label{fig:eval_deployment}
\end{figure}

Following the positive results of the validation in the simulated environment,  the agent was deployed in the physical MuVacAS protoype. A sequence of progressively complex experiments was conducted, ranging from maintaining fixed pressure targets to tracking dynamically varying setpoints over extended periods. During deployment, built-in safety protocols were active, including automatic shutdown upon exceeding pressure limits and the ability for a human operator to assume manual control. All agent actions and observations were logged in real time, providing a comprehensive dataset for evaluating the performance and stability of the autonomous control policy under real-world conditions (Figure~\ref{fig:eval_deployment} and Appendix~\ref{ap:drl_eval_real}). Notably, due to time constraints, the experimental setup featured a different argon injection location than the one encountered during training. Despite this change, the agent successfully reached and maintained all desired pressure configurations, though it required a larger number of control actions to do so.

\section{Conclusions and future work}

The deployment of the DRL agent on the physical MuVacAS prototype marks a significant step toward autonomous control of complex, non-linear subsystems in advanced scientific facilities. Trained on data from a prototype configuration in which argon injection occurred at the collision chamber, the agent was later deployed with the injector positioned further away (consistent with the geometry foreseen for the final accelerator design) to reduce irradiation exposure. This modification altered the pressure dynamics and introduced a distribution shift between the training and deployment conditions. Despite this, the agent successfully regulated argon injection to reach and maintain target pressures within operational bounds, even under dynamic scenarios not encountered during training. In contrast, proportional–integral–derivative (PID) controllers tuned under the previous configuration were unable to achieve satisfactory performance. The DRL agent’s adaptive and robust behaviour thus validates the feasibility of the learned policy and demonstrates the superior generalization capabilities of DRL compared with traditional control strategies. Broadly, these results provide a compelling proof-of-concept for AI-driven regulation of critical vacuum and gas injection systems in nuclear fusion facilities such as IFMIF-DONES, highlighting the potential of DRL to manage complex systems while reducing the operational load on human operators.

A key enabling factor in this success is the synergy between DLSMs and DRL. By providing a high-fidelity yet computationally efficient digital twin of the argon injection system, the FNO-based surrogate allowed the agent to be trained orders of magnitude faster than would be feasible on the physical system or with conventional simulators (e.g., Molflow~\citep{Kersevan:2694236-molflow}), where each control step requires approximately one second. This underscores the critical role of DLSMs in making DRL practical for real-world scientific applications. Beyond serving as safe and accelerated training environments, DLSMs can be used to construct digital twins of accelerator subsystems thanks to their near real-time inference capability, and they further open the door to advanced optimization strategies such as gradient-based methods due to their differentiability~\citep{mdpi-ddlsms}.

Although the main objective of this study was to demonstrate the synergy between DLSMs and DRL for autonomous pressure control, it is acknowledged that alternative learning paradigms could also address this problem effectively given the available data. Offline reinforcement learning methods, which derive optimal control policies directly from pre-collected datasets without further interaction with the environment, and imitation learning approaches, which learn control strategies from expert demonstrations, represent promising complementary directions. Future work will explore the implementation and benchmarking of these techniques against the current DLSM–DRL framework. 

Future iterations of the system will also aim to restrict the agent’s observation space to pressure readings obtained exclusively outside the collision chamber, thereby improving its suitability for deployment in the final accelerator configuration. In parallel, several additional enhancements are foreseen to further improve the agent’s performance. Fine-tuning the pre-trained policy on the real machine could help close the remaining sim-to-real gap~\citep{DBLP:journals-sim2real} and enhance robustness, particularly in regimes the agent has not previously encountered, such as modified injection setups. Moreover, the surrogate model’s accuracy can be improved by incorporating more representative transition states, as Gaussian random fields do not capture static or slowly varying states. Finally, planned future experiments will include the original injection to benchmark the agent against traditional controllers performance.

\begin{ack}
We wish to acknowledge and thank the other participants of DONES-FLUX project, the IFMIF-DONES consortium as well as the University of Granada for their role played in the development of this work. The project DONES-FLUX, with file number MIP-20221017, has been subsidised by the CDTI—Centro de Desarrollo Tecnológico Industrial through the call for proposals of the “Misiones CDTI” programme for the year 2022 and it is supported by the Ministry of Science and Innovation of Spain.
\end{ack}

\newpage
\bibliography{references} 

@article{ifmif-dones-status,
    author = {Bernardi, Davide and Ibarra, Angel and Arbeiter, Frederik and Arranz, F. and Cappelli, Mauro and Cara, Philippe and Castellanos, J. and Dzitko, Hervé and García, A. and Gutiérrez, Jandri and Królas, W. and Martin-Fuertes, Francisco and Micciché, G. and Muñoz Roldan, Antonio and Nitti, F. and Pinna, Tonio and Podadera, Ivan and Pons, J. and Qiu, Yuefeng and Román, Raquel},
    year = {2022},
    month = {10},
    pages = {24},
    title = {The IFMIF-DONES Project: Design Status and Main Achievements Within the EUROfusion FP8 Work Programme},
    volume = {41},
    journal = {Journal of Fusion Energy},
    doi = {10.1007/s10894-022-00337-5}
}

@inproceedings{sabogal:ipac2023-thpa156-muvacas,
    author = {Sabogal, A. and Others, B. and Others, C.},
    title = {Multipurpose Vacuum Accident Scenarios (MuVacAS) prototype for the IFMIF-DONES linear accelerator},
    booktitle = {Proc. IPAC'23},
    pages = {4324-4327},
    paper = {THPA156},
    venue = {Venice, Italy},
    series = {IPAC'23 - 14th International Particle Accelerator Conference},
    number = {14},
    publisher = {JACoW Publishing, Geneva, Switzerland},
    month = {05},
    year = {2023},
    issn = {2673-5490},
    isbn = {978-3-95450-231-8},
    doi = {10.18429/JACoW-IPAC2023-THPA156},
    url = {https://indico.jacow.org/event/41/contributions/2360},
    language = {English}
}

@article{Degrave2022,
  author    = {Jonas Degrave and Federico Felici and Jonas Buchli and Michael Neunert and Brendan Tracey and Francesco Carpanese and Timo Ewalds and Roland Hafner and Abbas Abdolmaleki and Diego de las Casas and Craig Donner and Leslie Fritz and Cristian Galperti and Andrea Huber and James Keeling and Maria Tsimpoukelli and Jackie Kay and Antoine Merle and Jean-Marc Moret and Seb Noury and Federico Pesamosca and David Pfau and Olivier Sauter and Cristiano Sommariva and Stefano Coda and Bruno Lepape and Martin Riedmiller and Olivier Bachem and Rupesh Srivastava and Timothy P. Lillicrap and Martin A. Riedmiller},
  title     = {Magnetic control of tokamak plasmas through deep reinforcement learning},
  journal   = {Nature},
  year      = {2022},
  volume    = {602},
  number    = {7897},
  pages     = {414--419},
  month     = feb,
  doi       = {10.1038/s41586-021-04301-9},
  url       = {https://www.nature.com/articles/s41586-021-04301-9},
  issn      = {1476-4687},
  publisher = {Springer Science and Business Media {LLC}}
}

@article{Lang_2011-GRF,
   title={Fast simulation of Gaussian random fields},
   volume={17},
   ISSN={1569-3961},
   url={http://dx.doi.org/10.1515/MCMA.2011.009},
   DOI={10.1515/mcma.2011.009},
   number={3},
   journal={Monte Carlo Methods and Applications},
   publisher={Walter de Gruyter GmbH},
   author={Lang, Annika and Potthoff, Jürgen},
   year={2011},
   month=jan }

@conference{osti_10193541-epics,
  author       = {Dalesio, L. R. and Hill, J. O. and Kraimer, M. and Lewis, S. and Murray, D. and Hunt, S. and Claussen, M. and Watson, W. and Dalesio, J.},
  title        = {The Experimental Physics and Industrial Control System architecture: Past, present, and future},
  booktitle    = {Proceedings of the International Conference on Accelerators and Large Experimental Physics Control Systems (ICALEPCS ’93)},
  year         = {1993},
  month        = {Nov},
  address      = {Berlin, Germany},
  organization = {Los Alamos National Laboratory},
  url          = {https://www.osti.gov/biblio/10193541}
}

@ARTICLE{2020arXiv201008895L-FNO,
       author = {{Li}, Zongyi and {Kovachki}, Nikola and {Azizzadenesheli}, Kamyar and {Liu}, Burigede and {Bhattacharya}, Kaushik and {Stuart}, Andrew and {Anandkumar}, Anima},
        title = "{Fourier Neural Operator for Parametric Partial Differential Equations}",
      journal = {arXiv e-prints},
         year = 2020,
        month = oct,
          eid = {arXiv:2010.08895},
        pages = {arXiv:2010.08895},
          doi = {10.48550/arXiv.2010.08895},
archivePrefix = {arXiv},
       eprint = {2010.08895},
 primaryClass = {cs.LG},
       adsurl = {https://ui.adsabs.harvard.edu/abs/2020arXiv201008895L}
}

@misc{towers2024gymnasiumstandardinterfacereinforcement,
      title={Gymnasium: A Standard Interface for Reinforcement Learning Environments}, 
      author={Mark Towers and Ariel Kwiatkowski and Jordan Terry and John U. Balis and Gianluca De Cola and Tristan Deleu and Manuel Goulão and Andreas Kallinteris and Markus Krimmel and Arjun KG and Rodrigo Perez-Vicente and Andrea Pierré and Sander Schulhoff and Jun Jet Tai and Hannah Tan and Omar G. Younis},
      year={2024},
      eprint={2407.17032},
      archivePrefix={arXiv},
      primaryClass={cs.LG},
      url={https://arxiv.org/abs/2407.17032}, 

}

@misc{schulman2017proximalpolicyoptimizationalgorithms,
      title={Proximal Policy Optimization Algorithms}, 
      author={John Schulman and Filip Wolski and Prafulla Dhariwal and Alec Radford and Oleg Klimov},
      year={2017},
      eprint={1707.06347},
      archivePrefix={arXiv},
      primaryClass={cs.LG},
      url={https://arxiv.org/abs/1707.06347}, 
}

@article{stable-baselines3,
  author  = {Antonin Raffin and Ashley Hill and Adam Gleave and Anssi Kanervisto and Maximilian Ernestus and Noah Dormann},
  title   = {Stable-Baselines3: Reliable Reinforcement Learning Implementations},
  journal = {Journal of Machine Learning Research},
  year    = {2021},
  volume  = {22},
  number  = {268},
  pages   = {1-8},
  url     = {http://jmlr.org/papers/v22/20-1364.html}
}

@article{DBLP:journals-sim2real,
  author       = {Wenshuai Zhao and
                  Jorge Pe{\~{n}}a Queralta and
                  Tomi Westerlund},
  title        = {Sim-to-Real Transfer in Deep Reinforcement Learning for Robotics:
                  a Survey},
  journal      = {CoRR},
  volume       = {abs/2009.13303},
  year         = {2020},
  url          = {https://arxiv.org/abs/2009.13303},
  eprinttype    = {arXiv},
  eprint       = {2009.13303},
  bibsource    = {dblp computer science bibliography, https://dblp.org}
}

@misc{Kersevan:2694236-molflow,
      author        = "Kersevan, Roberto and Ady, Marton",
      title         = "{Recent developments of Monte-Carlo codes Molflow+ and
                       Synrad+}",
      reportNumber  = "CERN-ACC-2019-173",
      pages         = "TUPMP037",
      year          = "2019",
      url           = "https://cds.cern.ch/record/2694236",
      doi           = "10.18429/JACoW-IPAC2019-TUPMP037",
}

@ARTICLE{mdpi-ddlsms,
       author = {{Gallardo}, Galo and {Rodr{\'\i}guez-Llorente}, Guillermo and {Magari{\~n}os Rodr{\'\i}guez}, Lucas and {Morant Navascu{\'e}s}, Rodrigo and {Khvatkin Petrovsky}, Nikita and {Lorenzo Ortega}, Rub{\'e}n and {G{\'o}mez-Espinosa Mart{\'\i}n}, Roberto},
        title = "{Differentiable Deep Learning Surrogate Models Applied to the Optimization of the IFMIF-DONES Facility}",
      journal = {Particles},
         year = 2025,
        month = feb,
       volume = {8},
       number = {1},
          eid = {21},
        pages = {21},
          doi = {10.3390/particles8010021},
       adsurl = {https://ui.adsabs.harvard.edu/abs/2025Parti...8...21G}
}

@inproceedings{
    rodriguezllorente2024applications-iclr,
    title={Applications of Fourier Neural Operators in the Ifmif-Dones Accelerator},
    author={Guillermo Rodr{\'\i}guez-Llorente and Galo Gallardo Romero and Roberto G{\'o}mez-Espinosa Mart{\'\i}n},
    booktitle={ICLR 2024 Workshop on AI4DifferentialEquations In Science},
    year={2024},
    url={https://openreview.net/forum?id=FL6ePnpBhB}
}

@article{wang2021surrogate,
  title        = {Surrogate model enabled deep reinforcement learning for hybrid energy community operation},
  author       = {Wang, Xiaodi and Liu, Youbo and Zhao, Junbo and Liu, Chang and Liu, Junyong and Yan, Jinyue},
  journal      = {Applied Energy},
  volume       = {289},
  year         = {2021},
  pages        = {116722},
  doi          = {10.1016/j.apenergy.2021.116722},
}
\bibliographystyle{plainnat} 

\newpage
\appendix

\section{Iterative evaluation of the surrogate model}
\label{ap:fno_eval}

\begin{figure}[H]
  \centering
  \includegraphics[width=1\linewidth]{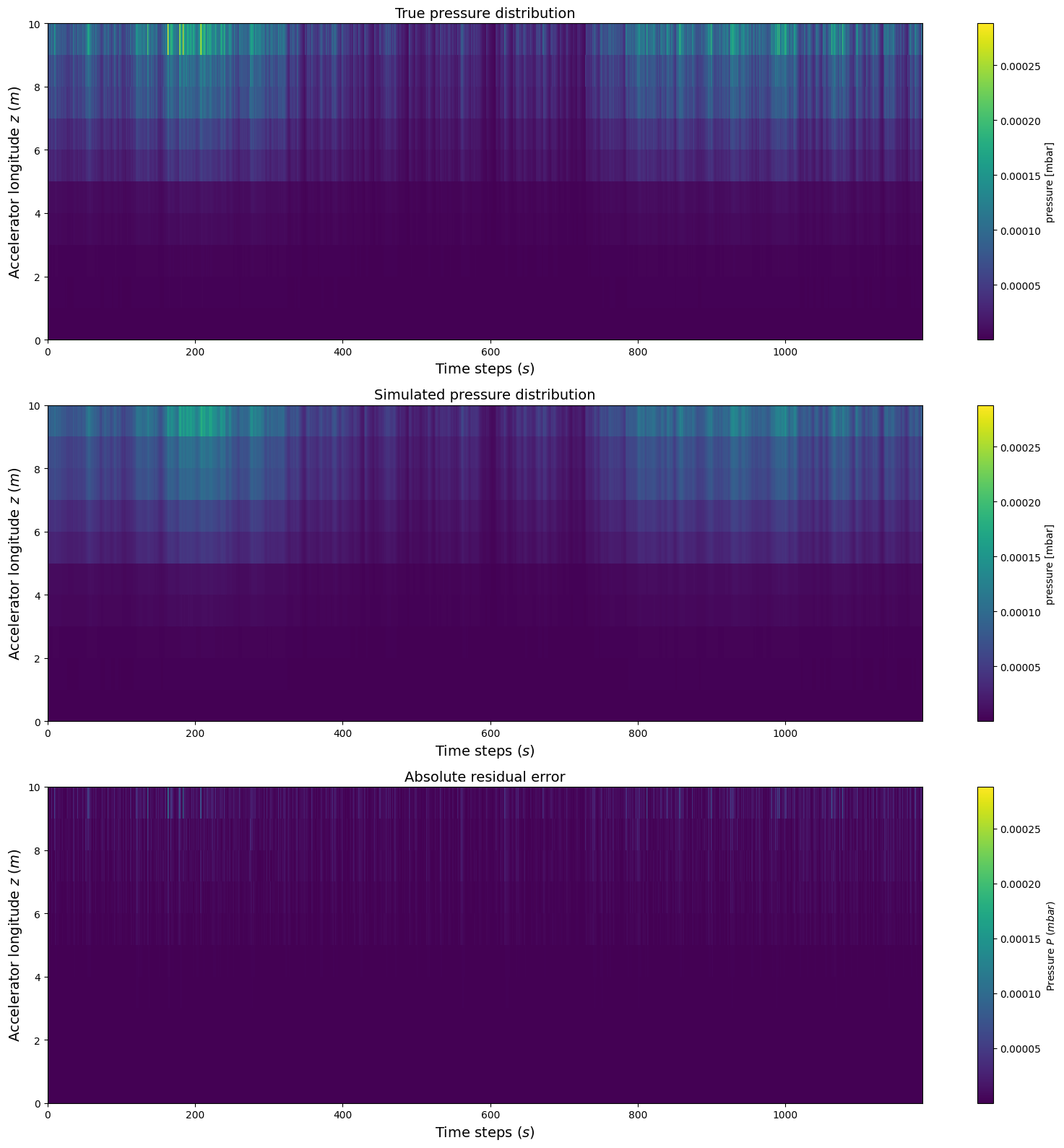}
  \caption{Iterative evaluation through time of the FNO surrogate model. It takes an initial pressure distribution along the accelerator longitude (in arbitrary units, each point represents a sensor) and different argon injections in each step.  The vertical axis in each plot represents $z$, the colorbar represents  the pressure $p$ and the horizontal axis is the time $t$. From top to bottom: the real pressure distributions from the test dataset, the simulated by the model pressure distributions and the residual errors.}
  \label{fig:eval_fno}
\end{figure}

\newpage
\section{Hyperparameters of the FNO surrogate model}
\label{ap:fno_hyper}

\begin{table}[ht]
  \caption{NVIDIA Modulus FNO architecture hyperparameters. Software versions: \texttt{nvidia-modulus} 0.6.0 and \texttt{nvidia-modulus.sym} 1.5.0, Ubuntu 22.04.4 LTS (Docker), and Torch 2.3.0.}
  \label{tab:fno_hparams}
  \centering
  \begin{tabular}{lll}
    \toprule
    Hyperparameter & Value & Description \\
    \midrule
    scheduler              & tf\_exponential\_lr & Learning rate schedule type \\
    optimizer              & adam               & Optimization algorithm \\
    loss                   & sum                & Loss aggregation method \\
    decoder.nr\_layers     & 1                  & Number of decoder layers \\
    decoder.layer\_size    & 256                & Hidden units per decoder layer \\
    fno.dimension          & 1                  & Spatial dimension of the problem \\
    fno.nr\_fno\_layers    & 6                  & Number of FNO layers \\
    fno.fno\_modes         & 12                 & Retained Fourier modes per layer \\
    scheduler.decay\_rate  & 0.95               & Exponential LR decay factor \\
    scheduler.decay\_steps & 1000               & Steps before each LR decay \\
    training.max\_steps    & 10000              & Total training iterations \\
    batch\_size.grid       & 32                 & Batch size for training data \\
    batch\_size.validation & 32                 & Batch size for validation data \\
    \bottomrule
  \end{tabular}
\end{table}

\section{Variables and hyperparamerers of the DRL problem}
\label{ap:drl_spaces}

\begin{table}[h]
  \caption{Observation and action space variables with their type in \texttt{Gymnasium}, ranges, and descriptions.}
  \label{tab:env_spaces}
  \centering
  \begin{tabular}{llll}
    \toprule
    Variable & Type     & Range           & Description \\
    \midrule
    $p_t$       & Box(1,)  & [$-6$, $-3$]    & Current pressure ($\log_{10}(\mathrm{mbar})$) at sensor 12 \\
    $p_{obj}$       & Box(1,)  & [$-6$, $-3$]    & Target pressure ($\log_{10}(\mathrm{mbar})$) \\
    $q_t$       & Box(1,)  & [0.0, 1.0]      & Argon injection rate (sccm, normalized) \\
    action  & Box(1,)  & [0.01, 0.9]     & Action value: new argon injection rate \\
    \bottomrule
  \end{tabular}
\end{table}

\begin{figure}[H]
  \centering
  \includegraphics[width=1\linewidth]{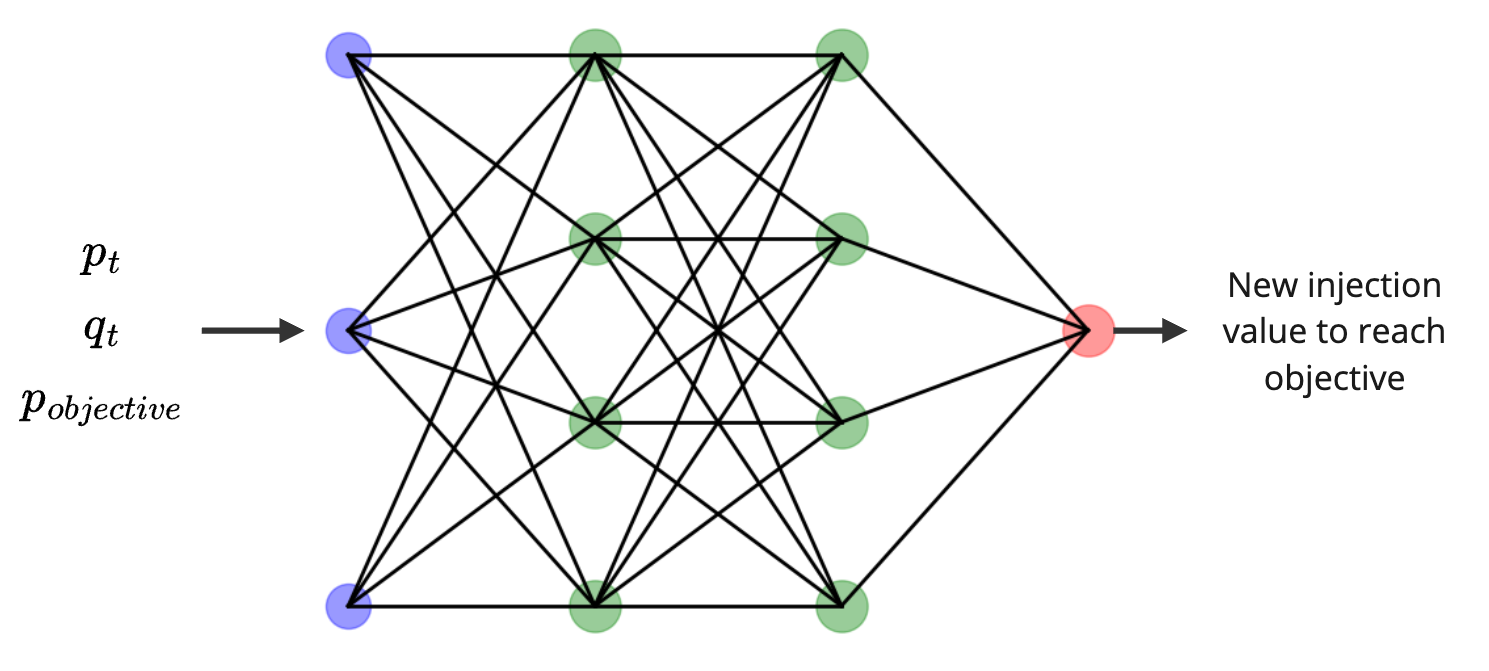}
  \caption{Policy neural network. Its inputs and output are described in Table \ref{tab:env_spaces}.}
  \label{fig:network_policy}
\end{figure}

\begin{table}[ht]
  \caption{PPO hyperparameters used with Stable-Baselines3. The agent training was stopped at 270k steps.}
  \label{tab:ppo_hparams}
  \centering
  \begin{tabular}{lll}
    \toprule
    Hyperparameter & Value & Description \\
    \midrule
    policy              & MultiInputPolicy & Policy network type (supports dict/obs inputs) \\
    learning\_rate      & 0.0008           & Step size for gradient descent \\
    n\_steps            & 2048             & Rollout steps per update \\
    batch\_size         & 64               & Minibatch size for each gradient update \\
    n\_epochs           & 10               & Number of epochs per update \\
    gamma               & 0.902            & Discount factor for rewards \\
    gae\_lambda         & 0.95             & GAE parameter for bias–variance trade-off \\
    clip\_range         & 0.2              & PPO clipping parameter \\
    clip\_range-vf      & null             & Clipping for value function (disabled) \\
    normalize\_advantage& true             & Advantage normalization enabled \\
    ent\_coef           & 0                & Entropy bonus coefficient \\
    vf\_coef            & 0.5              & Value function loss coefficient \\
    max\_grad\_norm     & 0.5              & Gradient clipping norm \\
    total\_timesteps    & 1,000,000        & Training budget in timesteps \\
    act\_function       & tanh             & Activation function in policy network \\
    layers              & [64, 64]         & Policy and value network architecture \\
    \bottomrule
  \end{tabular}
\end{table}

\section{Agent evaluation within the digital environment}
\label{ap:drl_eval_sim}

\begin{figure}[H]
  \centering
  \includegraphics[width=1\linewidth]{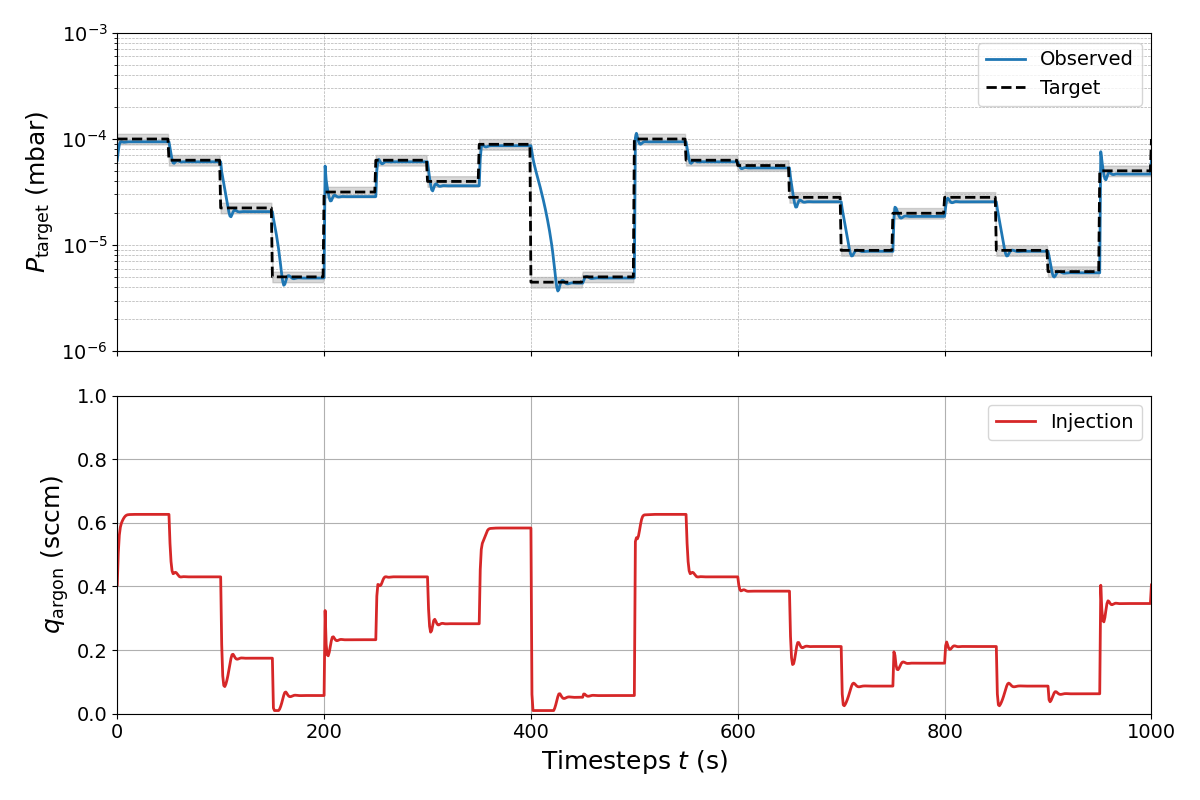}
  \caption{Evaluation of the DRL agent  on a different simulated environment. Top: observed (achieved by the agent) and objective (reference) pressures. Bottom: argon injection values.}
  \label{fig:eval2}
\end{figure}

\begin{figure}[H]
  \centering
  \includegraphics[width=1\linewidth]{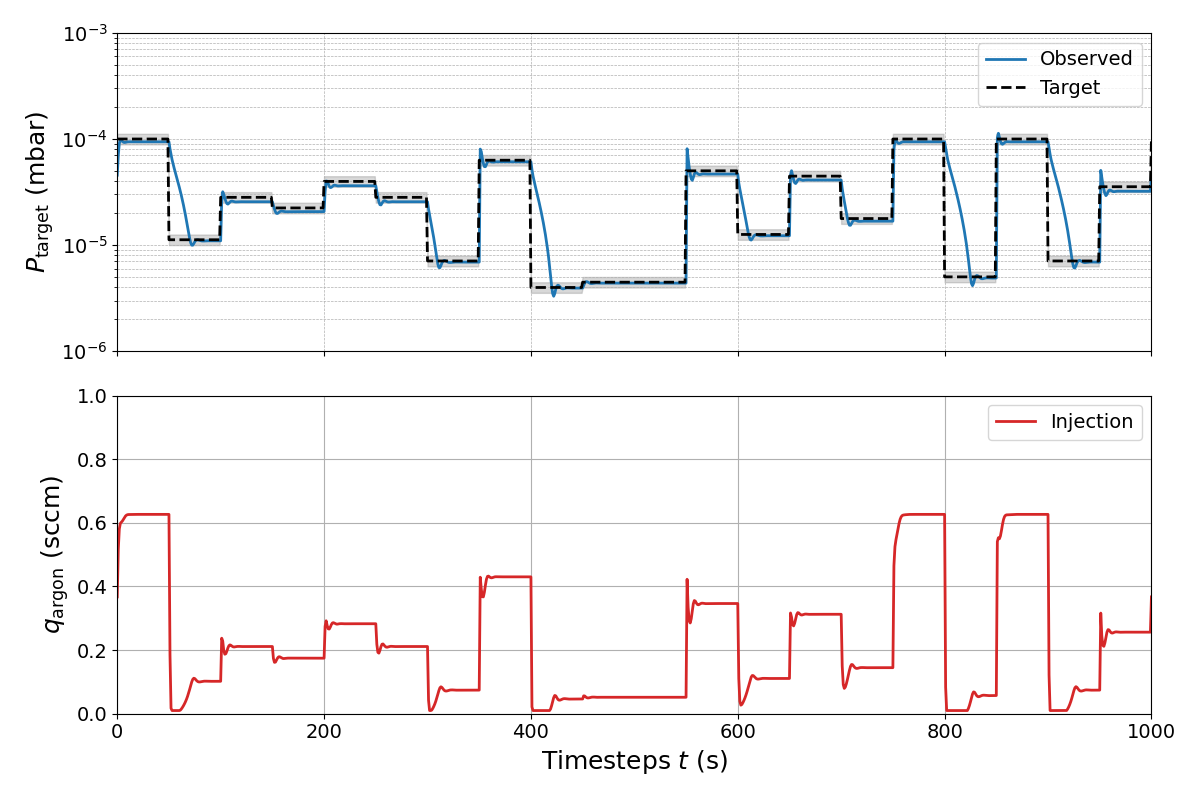}
  \caption{Evaluation of the DRL agent  on a different simulated environment. Top: observed (achieved by the agent) and objective (reference) pressures. Bottom: argon injection values.}
  \label{fig:eval3}
\end{figure}

\section{Agent evaluation within the real prototype}
\label{ap:drl_eval_real}

\begin{figure}[H]
  \centering
  \includegraphics[width=1\linewidth]{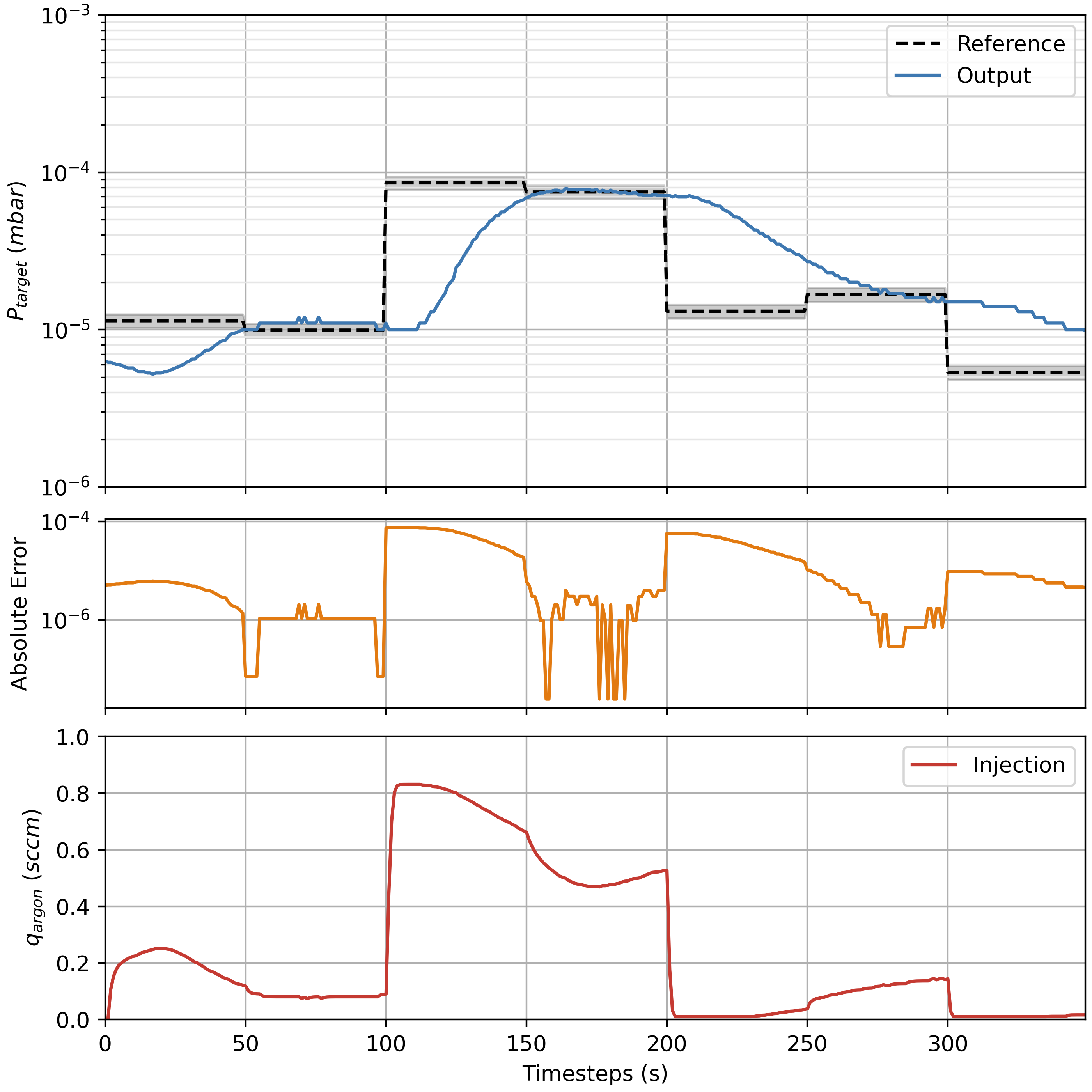}
  \caption{Evaluation of the DRL agent on the real prototype. Top: measured and objective pressures. Middle: absolute error between objective and achieved pressures. Bottom: argon injection rates. In this test, the transition interval between target pressure changes is set to 50 seconds. Because of the injector’s position in the system, the agent cannot fully reach the target within this time frame. Extending the transition interval, however, enables the agent to achieve the desired pressure.}
  \label{fig:eval4}
\end{figure}

\begin{figure}[H]
  \centering
  \includegraphics[width=1\linewidth]{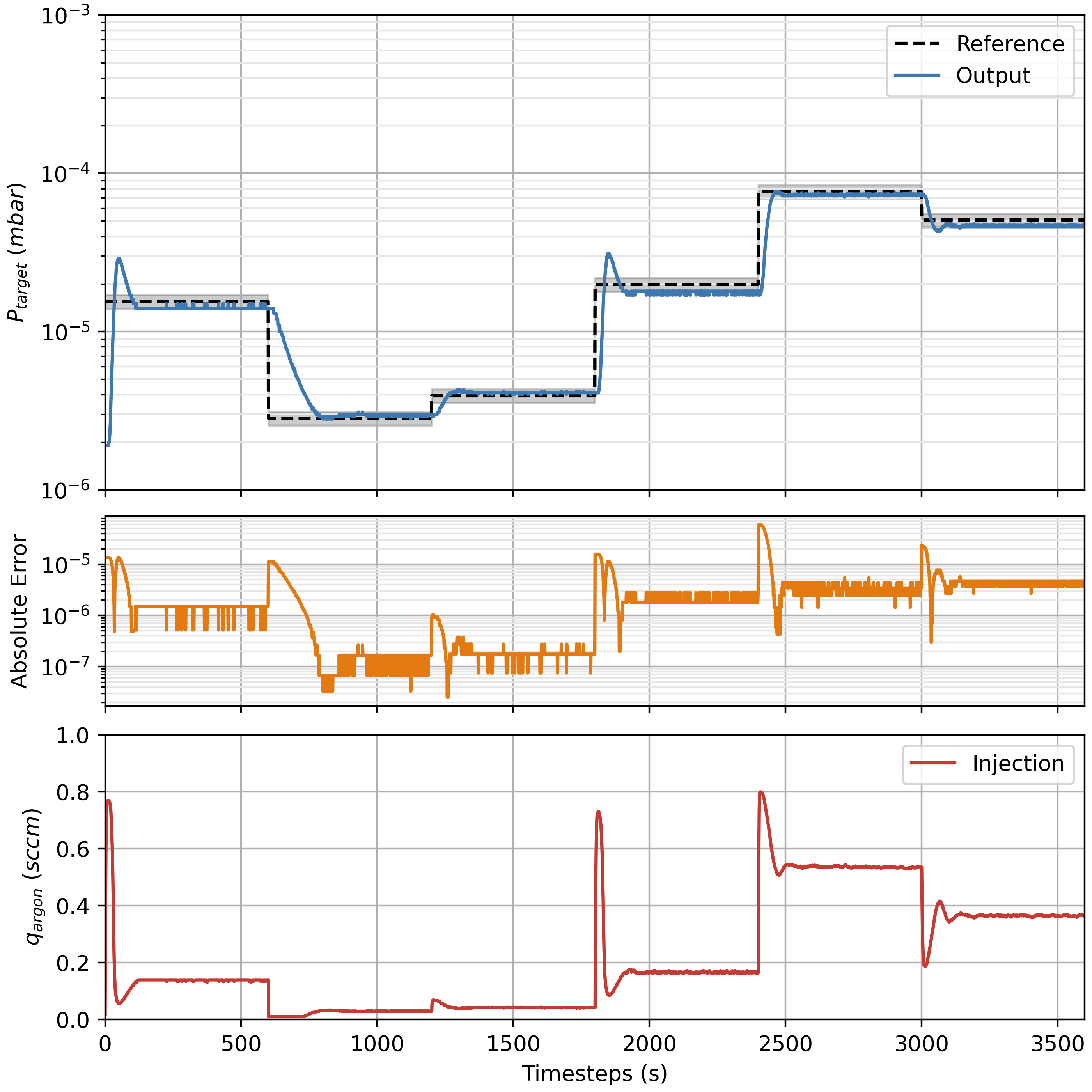}
  \caption{Evaluation of the DRL agent on the real prototype. Top: observed and objective pressures. Bottom: argon injection values.}
  \label{fig:eval5}
\end{figure}


\end{document}